\documentclass[pdflatex,sn-mathphys-num]{sn-jnl}

\usepackage{graphicx}%
\usepackage{subcaption} 
\usepackage{multirow}%
\usepackage{amsmath,amssymb,amsfonts}%
\usepackage{amsthm}%
\usepackage{mathrsfs}%
\usepackage[title]{appendix}%
\usepackage{xcolor}%
\usepackage{textcomp}%
\usepackage{manyfoot}%
\usepackage{booktabs}%
\usepackage{algorithm}%
\usepackage{algorithmicx}%
\usepackage{algpseudocode}%
\usepackage{listings}%

\theoremstyle{thmstyleone}%
\theoremstyle{thmstyletwo}%

\theoremstyle{thmstylethree}%

\begin{document}

\title[Workload composition smooths aggregate power demand while sustaining short-horizon ramps in AI data centers]{Workload composition smooths aggregate power demand while sustaining short-horizon ramps in AI data centers}

\author[1,2]{\fnm{Subir} \sur{Majumder}}\email{subir-em@ieee.org}

\author[1]{\fnm{Minlan} \sur{Yu}}\email{minlanyu@seas.harvard.edu}

\author[1]{\fnm{Le} \sur{Xie}}\email{xie@seas.harvard.edu}

\affil*[1]{\orgdiv{School of Engineering and Applied Sciences}, \orgname{Harvard University}, \orgaddress{\street{150 Western Ave}, \city{Allston}, \postcode{02134}, \state{Massachusetts}, \country{USA}}}

\affil[2]{\orgdiv{Department of Electrical and Computer Engineering}, \orgname{Texas A\&M University}, \orgaddress{\street{188 Bizzell St}, \city{College Station}, \postcode{77843}, \state{Texas}, \country{USA}}}

\abstract{Artificial intelligence (AI) is driving rapid growth in electricity demand, yet the grid-facing power dynamics of AI data centers remain poorly understood. Here we show that, in shared-GPU systems, the composition of batch and inference workloads decouples aggregate power variability from short-horizon ramping. As the inference share rises, variability becomes U-shaped, whereas ramping becomes hump-shaped, particularly under higher loading. The magnitude and turning points of these patterns also depend on system loading. Using a trace-calibrated framework linking workload arrivals, queueing, scheduling, and GPU power, we show that the underlying mechanism is asymmetric. At intermediate workload mixes, queued batch jobs fill capacity left idle by fluctuating inference demand, reducing aggregate power variability. However, short-horizon ramping remains elevated because inference-side fluctuations propagate more directly into realized power. AI data centers should therefore be understood as dynamic systems whose workload composition shapes their grid impact.}


\maketitle

\section{Introduction}\label{sec1}

Artificial intelligence (AI) is rapidly becoming a major source of electricity demand. As deployment expands, AI data centers are expected to account for a growing share of power-system load in the United States and globally \cite{lbnl2024datacenters, IEA2024}. Yet their grid impact cannot be inferred from annual energy use alone. Power systems respond to demand over minutes to hours, so what matters operationally is not only how much electricity AI data centers consume, but also how quickly their loads rise, fall, and fluctuate. These short-horizon dynamics affect reliability, reserve requirements, infrastructure planning, emissions accounting, and the value of flexibility \cite{weron2014electricity, stenclik2021redefining, nerc2010flexibility, gattaciecca2020identifying, GHGProtocolScope2_2023, mural2026aigrid}. Most estimates of AI electricity demand, however, are based on aggregate GPU counts or annual energy use \cite{lee2024datacenter, masanet2020recalibrating}, and planning studies often represent data centers as exogenous hourly loads with stylized flexibility limits \cite{chen2021incentive, riepin2025spatio, acun2023carbon}. Such approaches miss the internal operating dynamics through which computational work arrives, waits for service, and is translated into realized power demand.

This gap matters increasingly because many AI data centers now colocate latency-sensitive inference and queue-managed batch workloads on shared-GPU systems \cite{qian2024alibaba, wang2025colocating, chen2024latency}. We refer to such systems as hybrid AI data centers. Batch and inference workloads differ sharply in temporal structure, resource requirements, and power trajectories \cite{10.1145/3620666.3651329, choukse2025power, stojkovic2024towards, chung2025ml}. Batch jobs can often wait in scheduler-managed queues \cite{verma2015large, yoo2003slurm, burns2016borg} and, in some settings, be interrupted and resumed through checkpointing \cite{eisenman2022check}. Online inference, by contrast, is latency-sensitive: production serving systems typically keep queues short \cite{stojkovic2025dynamollm} and sustain throughput through mechanisms such as continuous batching \cite{kwon2023efficient} or rerouting under saturation \cite{zhong2024distserve}. As a result, colocating batch and inference on shared-GPU systems may produce power dynamics that cannot be inferred from studying either workload in isolation.

Here we show that hybrid AI data centers do not behave as weighted averages of batch-only and inference-only systems. Instead, workload composition changes grid-facing power demand nonlinearly: aggregate variability becomes U-shaped as inference share rises, whereas short-horizon ramping becomes hump-shaped. Hybrid systems can therefore be smoother in aggregate than the batch-only and inference-only extremes while remaining harder for the grid to follow on short timescales. This non-additive behavior arises because batch and inference workloads propagate fluctuations into power differently. Batch workloads buffer burstiness through queues, whereas latency-sensitive inference transmits it more directly into realized power, with only partial smoothing from capacity saturation and longer request residence times. When the two workloads share GPUs, queued batch work can fill residual-capacity valleys left by fluctuating inference demand, reducing aggregate variability without comparably reducing short-horizon ramping. AI data centers should therefore be understood not simply as large electric loads but as dynamic systems whose internal workload interactions shape grid-facing demand. Variability and ramping are distinct grid-facing properties of AI data centers and cannot be inferred from annual energy use alone.

\section{Hybrid workload composition reshapes variability and ramping in AI data centers} \label{sec2}

\begin{figure}[!htbp]
\centering
\makebox[\textwidth][c]{
    \includegraphics[width=1.35\textwidth]{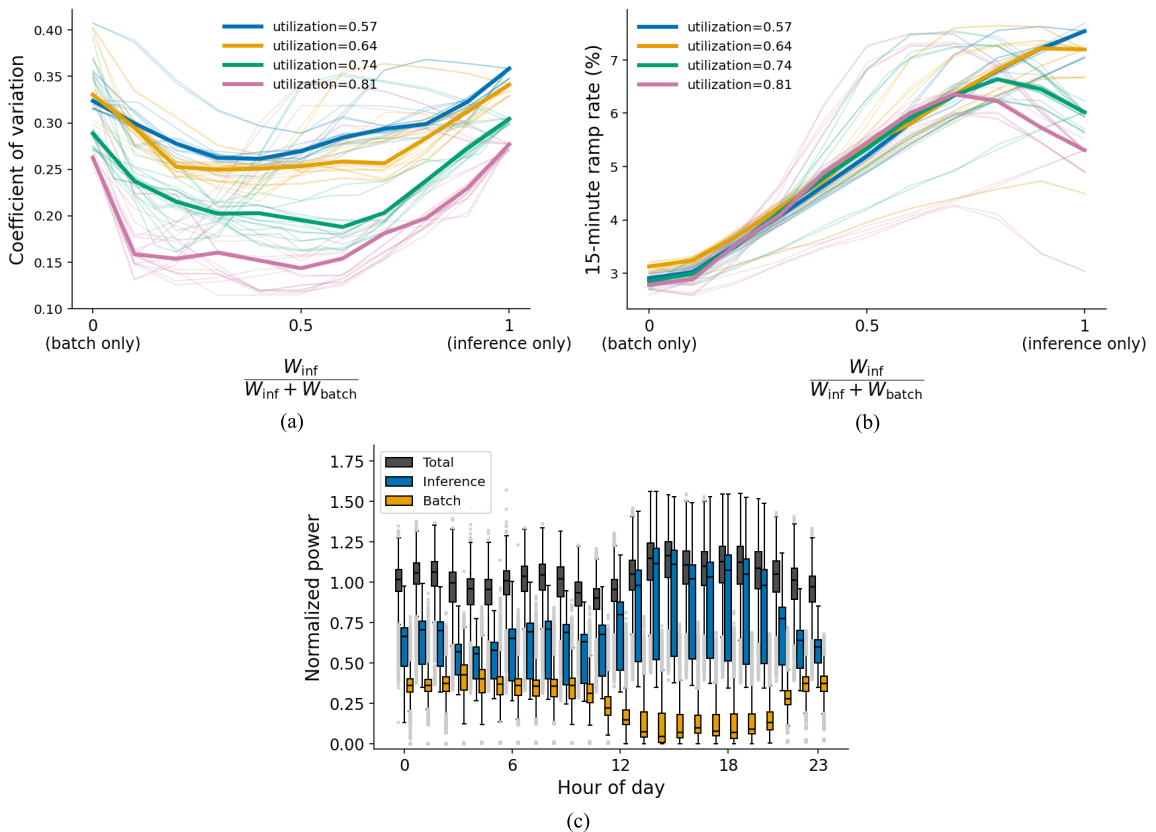}
}
\caption{Hybrid AI data center power varies nonlinearly with workload composition. (a) Coefficient of variation (COV) of total power versus the inference share of total offered load across utilization levels.
(b) Median normalized 15-minute ramp rate versus the inference share of total offered load for the same utilization levels.
(c) Hourly distributions of normalized total power in a representative hybrid batch--inference configuration, decomposed into inference and batch contributions. In (a) and (b), faint curves show individual configurations and bold curves show the median within each utilization level.
}\label{fig1}
\end{figure}

Hybrid batch--inference systems exhibit nonlinear load dynamics that cannot be inferred by interpolating between the batch-only and inference-only extremes. We find that hybrid systems are often less variable than either batch-only or inference-only systems, producing a U-shaped relationship between aggregate variability and inference share (Figure~\ref{fig1}(a); Supplementary Figure~20). Short-horizon ramping behaves differently. At higher utilization, the median short-horizon ramp rate becomes hump-shaped rather than monotonic as inference share rises (Figure~\ref{fig1}(b); Supplementary Figures~20 and~25). Thus, aggregate smoothness and short-horizon ramping decouple in shared-GPU AI data centers.

We show this using a trace-calibrated model of a hybrid AI data center in which batch jobs and inference requests share a common GPU pool. The model combines three empirical inputs: 98{,}000 batch jobs with per-job multi-GPU power profiles from MIT's high-performance computing facility \cite{samsi2021supercloud}, 44 million Azure LLM code and conversation requests \cite{stojkovic2025dynamollm}, and serving measurements that map inference token throughput to GPU demand and power \cite{chung2025ml}. In the model, latency-sensitive inference receives priority access to GPUs, while batch jobs use the remaining capacity \cite{wang2025colocating, chen2024latency}. To isolate the role of workload composition, we hold GPU count and utilization fixed while varying only the share of total work assigned to batch and inference, and we repeat this comparison across a range of scenarios (Supplementary Note~2, Supplementary Table~6; Methods). For each simulation, power profiles are normalized so that the comparison reflects workload interaction rather than system scale.

The reduction in variability at intermediate workload mixes arises mainly because batch jobs can wait. When inference demand is high, it takes GPU capacity immediately. Batch jobs, by contrast, can remain in the queue and start when inference leaves GPUs idle. This allows queued batch work to fill some of the gaps created by fluctuating inference demand, smoothing total GPU use and therefore aggregate power demand. The supplementary diagnostics support this interpretation. Batch GPU use increases when more capacity is left over after inference, and aggregate power closely follows total GPU use (Supplementary Figures~22--24). This smoothing effect becomes more important as the system gets busier and spare capacity becomes scarcer, which helps explain why the U-shape becomes more pronounced at higher utilization (Figure~\ref{fig1}(a); Supplementary Figure~20). How much smoothing batch jobs can provide also depends on how flexible they are. If batch jobs can be paused and resumed more frequently, they can respond more easily to changing leftover capacity. If checkpoint intervals are longer, batch work becomes less responsive, the minimum in variability shifts to lower inference shares, and variability rises sooner as inference becomes more dominant (Methods; Supplementary Figure~21). Partial cancellation between batch- and inference-driven fluctuations provides an additional source of smoothing (Supplementary Figure~27).

By contrast, batch- and inference-driven ramps offset one another only partially. As inference becomes a larger share of the workload, inference-driven ramps become more important because inference demand is transmitted more directly into realized power. At very high inference shares, however, this contribution flattens and can even decline, especially when the system is already heavily loaded, consistent with capacity saturation (Supplementary Figure~28). At the same time, the batch contribution to ramping shrinks because batch makes up a smaller share of the active workload. In hybrid systems, batch activity offsets some inference-driven fluctuations, so the realized system-wide ramp rate is usually smaller than the sum of the separate batch and inference ramp rates, but the offset is incomplete (Supplementary Figure~27). The resulting balance between growing inference-driven ramps, shrinking batch-driven ramps, incomplete offset, and saturation at high inference shares produces the hump-shaped pattern. This pattern also depends on utilization. At higher utilization, the batch contribution to ramping falls more quickly as inference share rises, especially when checkpoint intervals are long and batch jobs are less able to re-enter service in response to changing available capacity. At the system level, longer checkpoint intervals are associated with larger ramp rates that peak at lower intermediate inference shares (Supplementary Figure~26).

The same nonlinearity is visible in the daily time pattern. Inference demand tends to rise from midday into the evening, while batch activity declines, so aggregate power is flatter across the day than either component alone. This reshaping remains qualitatively similar across workload compositions and under alternative scheduling assumptions (Figure~\ref{fig1}(c); Supplementary Figures~17--19). Shared-GPU AI data centers can therefore become smoother in aggregate even while producing larger short-horizon ramps on the grid.

\section{Queue-buffered batch dynamics smooth hybrid power demand}

The hybrid results suggest that batch work reduces overall power variability because queued jobs can shift in time and use GPU capacity that would otherwise sit idle when inference demand fluctuates. To isolate that mechanism, we examine batch-only systems over a range of scenarios that vary utilization, checkpoint length, and the timing of job arrivals, while holding total GPU capacity fixed within each experiment set (Methods; Supplementary Table~7). The central question is how strongly short-horizon fluctuations in batch-job arrivals are reflected in realized execution and power.

\begin{figure}[!htbp]
\centering

\makebox[\textwidth][c]{
    \includegraphics[width=1.2\textwidth]{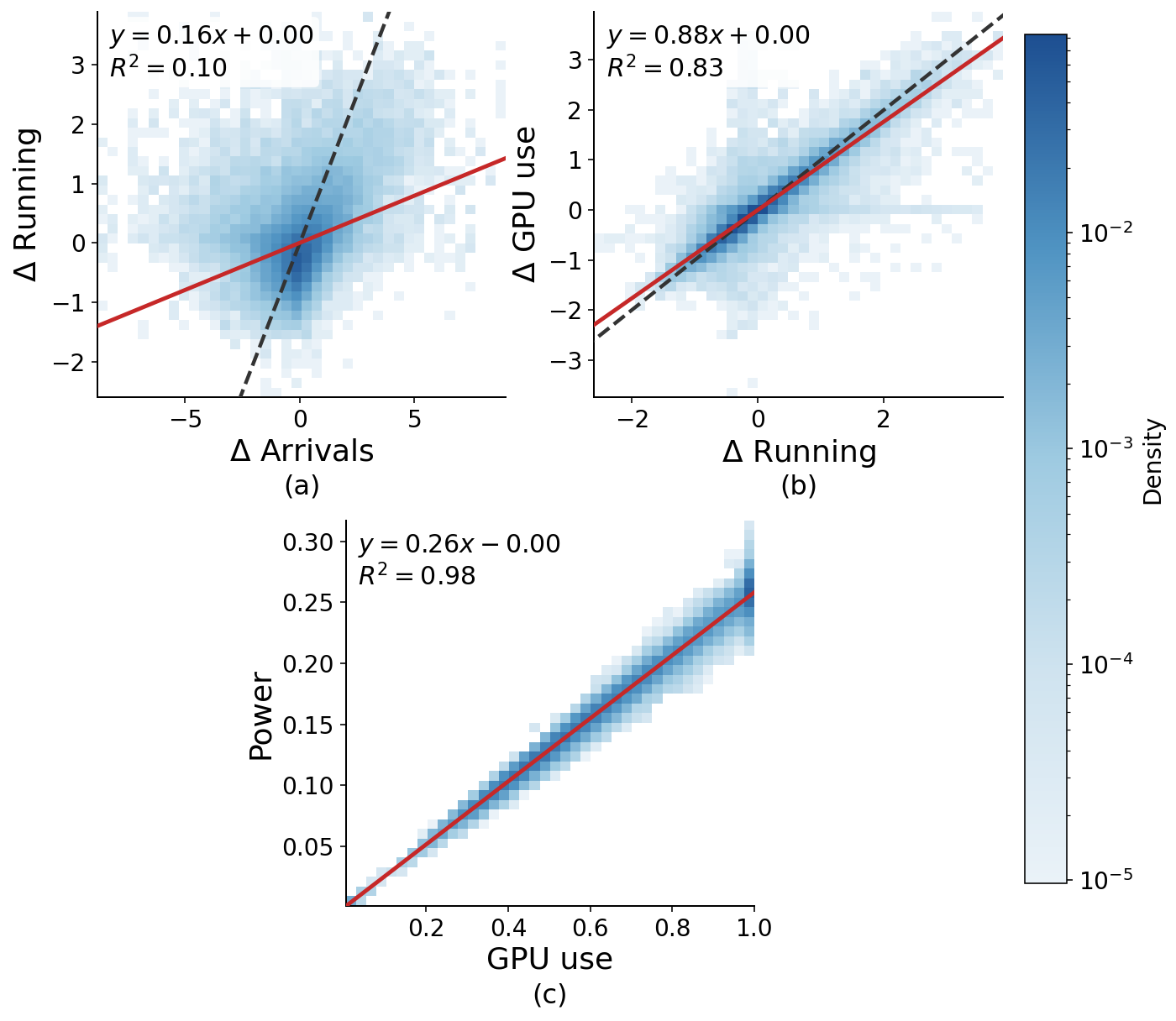}
}
\caption{Queue-mediated smoothing in batch-only systems.
(a) Relationship between 4-hour changes in standardized batch-job arrivals and running jobs. 
(b) Relationship between 4-hour changes in standardized running jobs and GPU use. 
(c) Relationship between normalized GPU use and normalized batch-system power. 
Each panel pools 4-hour observations across batch-only simulation runs. In (a) and (b), variables are standardized within run and first-differenced across consecutive 4-hour intervals; in (c), GPU use is normalized by the configuration-specific GPU count and power by the corresponding reference GPU power (0.7~kW per GPU). Blue shading shows pooled observation density, red lines show fitted ordinary least-squares relationships, and dashed lines in (a) and (b) mark the $45^\circ$ reference.}\label{fig:batch}
\end{figure}

We find that the connection between arrivals and realized power is weak initially and strong only later in the process. The median daily batch-power profile shows only modest recurring time-of-day structure (Supplementary Table~7; Supplementary Figure~29), but the more important result is that short-horizon changes in job arrivals are only loosely related to changes in the number of jobs that are actually running (Figure~\ref{fig:batch}; Supplementary Figures~30--32). Much of the burstiness in arrivals is therefore absorbed by the queue before it reaches active execution. This transmission is somewhat stronger at lower utilization and when batch jobs are more flexible, because the scheduler has more freedom to release queued work rather than continue to hold it back, but it remains weak overall. Once jobs leave the queue, however, the mapping becomes much tighter. Changes in the number of running jobs are closely associated with realized GPU use, with little variation across utilization and checkpoint settings. Realized GPU use is then closely and approximately linearly associated with realized batch power. The batch-side mechanism is therefore sequential: queues absorb much of the arrival-side burstiness, scheduling determines when buffered work enters service, and realized GPU use translates those execution decisions into power demand. In the hybrid system, this same queue-mediated flexibility allows batch work to fill capacity left unused by fluctuating inference demand, thereby smoothing aggregate power.

Serving conditions determine how strongly batch can play this smoothing role. In the batch-only system, overall variability is governed mainly by utilization, with only modest influence from checkpoint length (Supplementary Figure~33). As utilization rises, both variability and the ramp rate decline, while per-job mean-delay increases. Shorter checkpoint lengths are associated with slightly lower variability and delay but somewhat larger ramp rates. In the hybrid system, these same factors determine how quickly queued batch work can re-enter service when inference leaves capacity available. This helps explain why the variability-reducing effect of batch depends on utilization and checkpoint length in the hybrid results, and why the batch contribution to ramping declines as utilization rises and batch-job flexibility falls.

\section{Inference dynamics sustain short-horizon ramping in hybrid systems}

The hybrid results show that short-horizon ramping can remain high even when overall variability falls at intermediate workload mixes. This asymmetry points to the inference side of the system. Unlike batch jobs, inference requests are latency-sensitive, so they are usually served with little waiting \cite{stojkovic2025dynamollm}. The main question is how strongly short-horizon fluctuations in inference demand are reflected in realized power. To isolate that mechanism, we examine inference-only systems over a range of scenarios that vary utilization, GPU capacity, maximum inference batch size, and verbosity scale (Methods; Supplementary Table~8).

\begin{figure}[!htbp]
\centering
\makebox[\textwidth][c]{
    \includegraphics[width=1.2\textwidth]{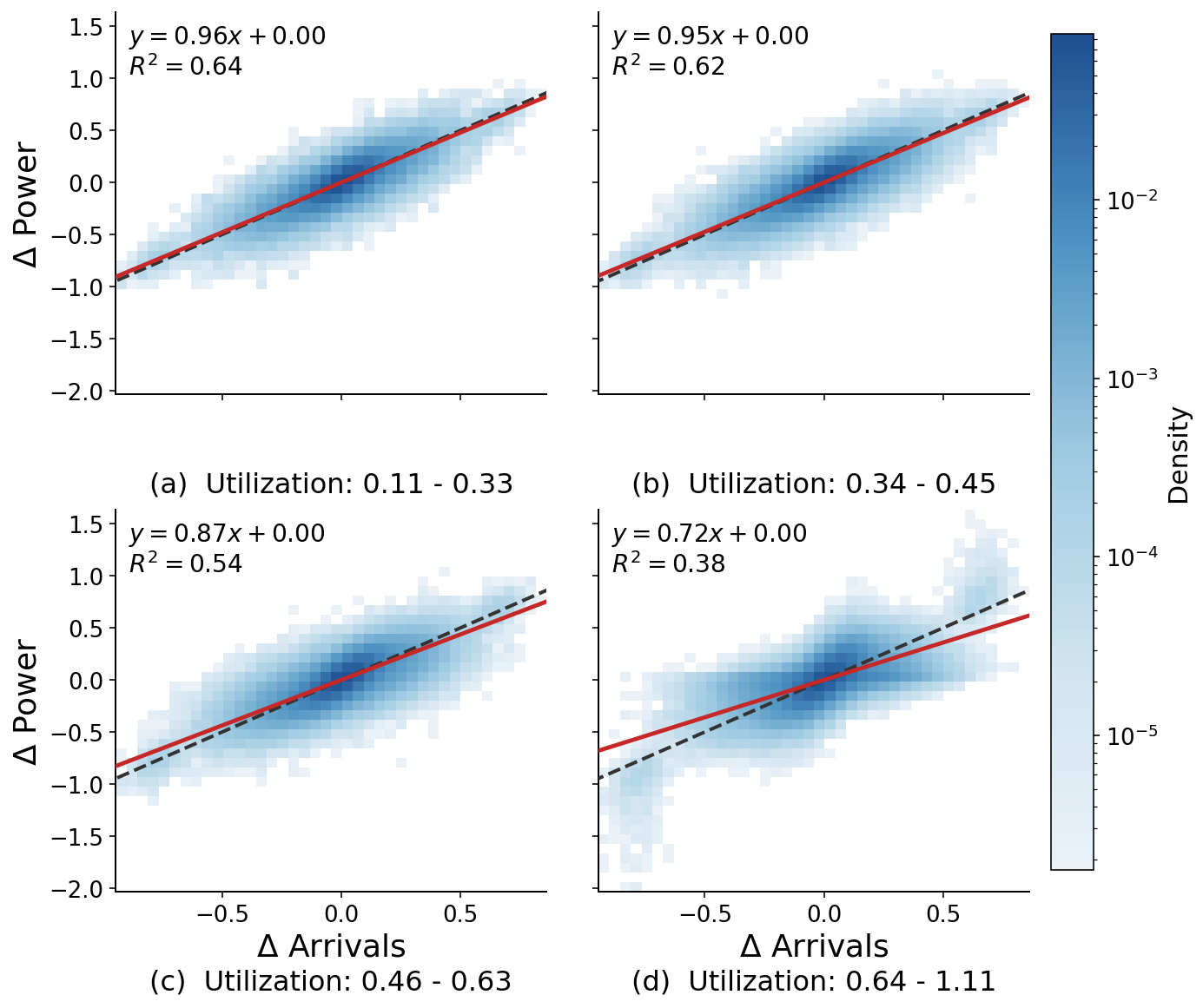}
}
\caption{Short-horizon changes in inference arrivals and power by utilization. 
Each panel pools 15-minute observations across inference-only simulation runs. Variables are standardized within run and first-differenced across consecutive 15-minute intervals. Panels pool standardized 15-minute changes across runs within utilization quartiles; the corresponding utilization intervals are reported in the panel labels. Colors indicate pooled observation density, the red line shows the pooled ordinary least-squares fit, and the dashed line denotes the $45^\circ$ reference.}\label{fig:inf}
\end{figure}

We find that fluctuations in inference demand are transmitted into realized power much more directly than in batch-only systems, although this transmission weakens as utilization rises and requests remain active for longer. Across utilization levels, changes in request arrivals are positively associated with changes in power, and this pattern remains qualitatively similar across serving-speed and verbosity settings (Figure~\ref{fig:inf}; Supplementary Figures~35--38). At high utilization, however, the relationship becomes flatter at large arrival changes: larger bursts in arrivals no longer produce proportionally larger changes in realized power. This flattening is consistent with capacity saturation. When the system is already heavily loaded, additional requests cannot increase power demand as quickly because available GPU capacity is limited. Request duration provides a second source of smoothing. Slower token generation keeps requests active for longer, which weakens the link between new arrivals and immediate changes in power. Larger maximum inference batch sizes can have a similar effect in vLLM-style inference systems by increasing time per output token \cite{kwon2023efficient}, and longer outputs likewise extend request duration. These duration-based smoothing effects are clearest at low utilization and become less pronounced as utilization rises. Even so, inference power retains a strong daily pattern across scenarios, with higher and more dispersed loads on weekdays than on weekends (Supplementary Table~9; Supplementary Figure~34).

These same mechanisms are reflected in the system-level variability and ramping metrics. Both the coefficient of variation and the ramp rate generally decline as utilization rises and verbosity increases (Supplementary Figures~39--41), but the change is larger for ramping than for variability, especially given the narrower ramping range observed in batch-only systems. This suggests that short-horizon ramping in hybrid systems is shaped more strongly by inference dynamics than by batch dynamics, and that the hump-shaped relationship emerges primarily from how inference-side fluctuations interact with utilization and saturation.

\section{Conclusion}\label{sec_conclusion}

AI data centers are not simply large electricity consumers whose grid impact can be inferred from annual energy use. Their time-resolved power demand depends on how batch and inference workloads share resources within the same computing system. When these workloads share GPUs, their aggregate power impact is not additive. Using a trace-calibrated shared-GPU framework, we show that workload composition, queueing, and scheduling jointly shape grid-facing power demand. Batch workloads buffer burstiness through queues, whereas inference workloads transmit fluctuations more directly into realized power, with only partial smoothing from saturation and longer request residence times. As a result, colocated systems exhibit nonlinear behavior: moderate amounts of queue-buffered batch work can smooth aggregate demand by filling residual-capacity valleys left by fluctuating inference demand, yet the same system can still impose substantial short-horizon ramps as inference becomes a larger share of the active workload. These effects also depend on utilization and workload-specific service parameters. Aggregate variability is U-shaped in workload composition, whereas short-horizon ramping becomes hump-shaped at higher utilization. The workload mix that minimizes variability therefore need not minimize ramping. Variability and ramping are distinct grid-facing properties of AI data centers.

Our analysis isolates workload-driven power dynamics rather than the full facility response. We do not model cooling systems, on-site storage, backup generation, or other control layers that may buffer or reshape load before it reaches the grid. Prior work suggests that some of this variability can be managed internally \cite{choukse2025power, ding2018emission, yu2016distributed}. Even so, those systems must still respond to the underlying signal created by workload arrivals and shared-resource contention. As AI deployment expands, credible grid planning and demand modeling will require moving beyond annual energy totals and static load assumptions to represent how digital-service dynamics translate into electrical demand.

\section{Methods}\label{sec_methods}

We model two GPU-intensive workload classes in an AI data center, namely batch training jobs and online inference requests. The two workloads share a common GPU pool under a shared-GPU colocation policy \cite{qian2024alibaba, wang2025colocating, chen2024latency}, in which latency-sensitive inference receives priority access to GPUs and batch jobs run on the residual capacity. The model combines three empirical inputs. First, batch arrivals, job characteristics, and per-job GPU power profiles are calibrated from the MIT SuperCloud trace \cite{samsi2021supercloud}. Second, inference arrivals and request characteristics are calibrated from anonymized request-level traces of large language model (LLM) services in Microsoft Azure \cite{stojkovic2025dynamollm}. Third, because the inference trace does not include direct power measurements, we augment it with independently reported serving-energy measurements across representative LLMs and serving configurations \cite{chung2025ml}, with archived values summarized in Supplementary Table~4. A schematic of the integrated model, together with an example time-resolved power profile, is shown in Supplementary Figure~16.

\subsection*{Workload arrival model}

\textit{Batch job arrivals.}
We model batch jobs as rigid jobs \cite{lublin2003workload} with fixed GPU requirements over the realized runtime of each job. To summarize heterogeneity in resource demand, we group jobs by unique $(\text{GPU count}, \text{time limit})$ combinations and apply agglomerative hierarchical clustering \cite{ward1963hierarchical} to $\log(\text{GPU count})$ and $\log(\text{GPU-hours})$, where $\text{GPU-hours}=\text{GPU count}\times \text{time limit (in hours)}$. This procedure yields three empirical resource groups---low, medium, and high---with distinct patterns over the measurement window (Supplementary Note~1 and Supplementary Figure~1).

For each group $g\in\{\text{low},\text{medium},\text{high}\}$, let $y_{g,t}$ denote the number of batch-job arrivals on day $t$. Because daily counts exhibit calendar structure and overdispersion, we model them using a negative-binomial (NB2) generalized linear model \cite{hilbe2011negative},
\[
y_{g,t} \sim \mathrm{NB}(\mu_{g,t}, \alpha), 
\qquad 
\log \mu_{g,t}
= \beta_{g,\text{wd/we}} 
+ \gamma_{\text{wom}(t)} 
+ \eta_{\text{week}(t)},
\]
where $\mu_{g,t}=\mathbb{E}[y_{g,t}]$ and $\mathrm{Var}(y_{g,t})=\mu_{g,t}+\alpha\mu_{g,t}^2$. Here $\beta_{g,\text{wd/we}}$ denotes group-specific weekday/weekend effects, $\gamma_{\text{wom}(t)}$ week-of-month effects, and $\eta_{\text{week}(t)}$ week fixed effects. We fit the model by maximum likelihood, with standard errors clustered by day (Supplementary Table~1). Simulation-based predictive checks show that the fitted model reproduces the observed mean and dispersion of daily arrivals reasonably well (Supplementary Figure~2). Because our objective is steady-state scenario generation rather than reconstruction of the realized growth path in the trace, synthetic arrivals are generated after averaging over the fitted week effects within each resource-group $\times$ day-type $\times$ week-of-month combination. This removes the realized week-to-week growth component while preserving recurrent calendar structure (Supplementary Figure~3).

Within-day arrival timing is modeled as compositional data on the simplex using a logistic-normal distribution \cite{aitchison1982statistical}. For day $t$ in group $g$, let $\mathbf{x}_{g,t}$ denote the vector of hourly arrival counts with total $N_{g,t}$, and define $\mathbf{w}_{g,t}=\mathbf{x}_{g,t}/N_{g,t}$. To avoid zero components, we apply a small uniform shrinkage step before normalization. We then apply an additive log-ratio transform to $\mathbf{w}_{g,t}$ using the hour with the largest mean probability as the reference, fit a multivariate normal distribution with diagonal covariance in transformed space, and invert the transform to generate synthetic hourly profiles. The fitted mean vector and diagonal covariance are calibrated to match two features of the observed trace: the pooled hourly arrival histogram and the standard deviation of the daily center of mass of arrivals (Supplementary Figure~4). Within each hour, arrivals are assumed uniform in time.

Because the observed batch trace plausibly reflects a single time zone, we also consider a multi-time-zone scenario variant of the batch-arrival model. In this variant, we simulate independent daily and intraday arrival processes across a fixed number of time zones, dividing total mean daily arrivals equally across zones so that expected total batch work is preserved. Each zone uses the same intraday profile shifted by a fixed clock-time offset. When re-aligned to a common reference time, the superposed process preserves expected total load while allowing fluctuations to smooth through temporal averaging across zones.

\textit{Inference request arrivals.}
The inference trace contains one week of request-level logs from a production LLM service, including timestamps, context tokens, generated tokens, and a code-assist versus conversation label. Over this short horizon, we do not model long-run growth, but we do model strong intraday variation and heterogeneity across request types. Following DynamoLLM \cite{stojkovic2025dynamollm}, we use context-token quantiles to define request groups. Because generated-token distributions are much less sensitive to context length for Code requests, we retain a single pooled Code group. Conversation requests are partitioned into four groups, Conversation Q1--Q4 (Supplementary Table~2).

For each request group $g\in\{\text{Code},\text{Conversation Q1},\dots,\text{Conversation Q4}\}$ and minute $t$, let $y_{g,t}$ denote the number of arrivals. We model minute-level arrivals as increments of a nonhomogeneous counting process with
\[
\mathbb{E}[y_{g,t}] = \mu_{g,t}, \qquad
\log \mu_{g,t} = \beta_g + f_g(\text{time-of-day}_t,\text{weekend}_t),
\]
where $f_g(\cdot)$ is a linear combination of 15-minute-of-day indicators, a weekend indicator, and their interactions (Supplementary Table~3). Residual variation remains overdispersed relative to a Poisson mean model, so we add a group-specific NB2 variance layer \cite{hilbe2011negative}, $\operatorname{Var}(y_{g,t}\mid g)=\mu_{g,t}+\alpha_g \mu_{g,t}^2$. We estimate $\alpha_g$ by the method of moments and then apply a single multiplicative calibration factor, shared across request groups, so that the overall empirical coverage of the 5--95\% predictive band for minute-level arrivals is close to nominal on the observed trace. The resulting fit is strongest for the Conversation groups and weaker for the Code group (Supplementary Figure~13). Arrival times are assumed uniform within each minute.

Because the trace does not identify the serving LLM model, we introduce seven representative LLM templates (Supplementary Table~4) as a modeling layer for scenario generation. For each request group, we simulate independent minute-level arrival processes across these templates, dividing total mean arrivals equally across templates so that expected group-level load is preserved. To avoid artificial smoothing when arrivals are split across templates, we rescale template-level overdispersion so that the superposed process preserves the variance of the original group-level arrival model.

\subsection*{Workload characteristics}

\textit{Batch workloads.}
For each batch-job resource group $c$, we model the joint distribution of time limit $tl$, GPU count $gpu$, and runtime as
\[
P\bigl(tl,gpu,\log(\mathrm{runtime}) \mid c\bigr)
=
P(tl \mid c)\,P(gpu \mid c,tl)\,P\bigl(\log(\mathrm{runtime}) \mid c,tl,gpu\bigr).
\]
Here, we use $tl$ as the primary stratification variable. We estimate $P(tl\mid c)$ and $P(gpu\mid c,tl)$ as multinomials with symmetric Dirichlet (add-$\alpha$) smoothing \cite{gelman1995bayesian} over the observed support within each group. For each observed leaf $(c,tl,gpu)$, we model $\log(\mathrm{runtime})$ using a hierarchical quantile function on a fixed 1\%--99\% probability grid. If a leaf has sufficient support, we use its empirical quantiles directly; otherwise we back off to the corresponding $(c,tl)$- or $c$-level distribution in Katz style \cite{katz1987estimation}. Synthetic jobs are generated by sampling $(tl,gpu)$ and then drawing runtime from the corresponding quantile curve, truncated at the requested time limit. Across Monte Carlo replications, the resulting joint distribution over $tl$, $gpu$, and $\log(\mathrm{runtime})$ matches the observed trace reasonably well (Supplementary Figures~5--6).

We model batch-job power using per-minute templates indexed by $(c,tl,gpu,\text{runtime-bin})$, where \textit{runtime-bin} denotes bins of $\log(\mathrm{runtime})$ within each $(c,tl,gpu)$ cell. Templates are estimated directly when a cell contains at least 194 jobs; otherwise we use a count-gated hard-backoff hierarchy in the spirit of Katz \cite{katz1987estimation},
\[
(c,tl,gpu,\text{runtime-bin}) \rightarrow (c,tl,gpu) \rightarrow (c,tl) \rightarrow c.
\] 
This count is a practical threshold motivated by variance-estimation precision. To construct a template, we resample each job's power trace to 1-minute mean power, normalize to a per-GPU basis, pool observations by minute index, and compute minute-wise means, standard deviations, and 5th/95th percentiles. Residual temporal dependence is represented by an AR(1) process whose coefficient is set to the median lag-1 autocorrelation across standardized job-level series within the template. To generate a job-level power trace, we combine the template's time-varying mean and standard deviation with the AR(1) residual process, clip simulated values to the empirical 5th--95th percentile band, and then rescale by the job's GPU allocation. Finally, we calibrate a single global noise-control factor, shared across templates, to align the synthetic aggregate GPU-power trace with the observed data-center GPU-power trace. The factor is chosen by empirical calibration against the observed aggregate 1-minute GPU-power trace to balance time-domain error and mean bias across Monte Carlo simulations. Under this single calibration, the synthetic aggregate trace is broadly consistent with the observed series in the time domain, in power spectral density, and in the marginal distribution of 1-minute power (Supplementary Figures~7--8).

\textit{Inference workloads.}
We model generated-token counts separately for Code requests and for the four Conversation groups. For Conversation Q1--Q4, generated-token counts are modeled on the discrete support $1,\dots,1200$. We first construct the pooled empirical histogram across all Conversation requests and smooth it by local kernel averaging. The resulting smoothed pooled histogram defines a Dirichlet prior with total pseudo-count mass $\tau$. For each Conversation group, the fitted probability mass function is the posterior mean, which shrinks sparse groups toward the pooled Conversation distribution. For Code requests, generated-token counts are modeled on the discrete support $1,\dots,5000$ using the pooled empirical histogram after the same smoothing step. We assess the fitted distributions against empirical complementary cumulative distribution functions (CCDFs) and group-specific histograms (Supplementary Figures~14--15). To study the effect of longer responses on compute demand, we introduce a global verbosity scale $s>0$ that approximately rescales generated lengths via
\[
F_{\mathrm{new}}(y) \approx F_{\mathrm{old}}(\lfloor y/s \rfloor).
\]

The inference trace does not identify the underlying LLM or hardware and contains no direct power measurements. We therefore augment request arrivals and generated-token counts with independently collected serving measurements for representative LLMs and serving configurations, summarized in Supplementary Table~4, obtained under continuous batching with vLLM \cite{kwon2023efficient}. For each representative LLM, we consider multiple serving configurations that differ in maximum inference batch size. Because larger maximum inference batch sizes are typically associated with higher per-request token-generation latency in vLLM-style systems \cite{10.1145/3620666.3651329}, we group these configurations into three serving-speed categories---F (fast), M (medium), and S (slow)---based on reported average time per output token in our dataset. For each LLM template and serving-speed setting, the serving measurements determine both the number of GPUs required per serving instance and an effective average power coefficient per active request. Because the reported per-request energy is averaged over execution and plausibly includes prefill overhead, and because our model does not resolve faster within-request power variation, this coefficient should be interpreted as an effective request-level decoding-window power rather than an instantaneous within-request power trajectory.

\subsection*{Scheduling and co-simulation}

\textit{Inference scheduling.}
The inference trace does not identify the production scheduling policy. We therefore model inference service using a stylized continuous-batching approximation \cite{kwon2023efficient}, consistent with the serving system used in the independently collected measurements. The arrival and workload-characteristics model generates request timestamps and generated-token counts for Code and Conversation Q1--Q4 traffic across seven representative LLM templates, each paired with a selected maximum inference batch size setting. Requests are then pooled by LLM template, because any instance of a given template can serve requests from any request group. Because online inference is latency-sensitive and production systems typically keep waiting times short \cite{stojkovic2025dynamollm}, we do not model a persistent internal inference queue. Instead, request token counts are mapped to effective service windows under a fixed-time service grid that provides a stylized approximation to short batching and dispatch delays under continuous batching. This approximation is used to estimate realized served concurrency and power, rather than end-user latency outcomes.

Let $\bar C_m(t)$ denote the minute-averaged number of concurrent active requests for template $m$ at minute $t$, $B_m$ the maximum inference batch size, and $g_m$ the number of GPUs required by one serving instance. When a finite serving budget is imposed, we allocate a template-specific GPU budget $G_m^{\max}$ in proportion to offered GPU-hours. We then define capped effective concurrency as
\[
\bar C_m^{\mathrm{cap}}(t)=\min \left\{ \bar C_m(t),\, \frac{B_m\,G_m^{\max}}{g_m} \right\},
\]
so that capped GPU use is
\[
G(t)=\sum_m g_m\left\lceil \frac{\bar C_m^{\mathrm{cap}}(t)}{B_m}\right\rceil.
\]
This cap provides a stylized representation of finite serving capacity. Excess demand is not modeled as an accumulating internal queue; instead, it is treated as unmet or externally managed request pressure, consistent with production mechanisms such as rate limiting \cite{sheng2024fairness}, rerouting \cite{zhong2024distserve}, or related control actions. With per-active-request power coefficient $\rho_m$, total inference power is
\[
P(t)=\sum_m \rho_m\,\bar C_m^{\mathrm{cap}}(t).
\]

\textit{Batch scheduling.}
Batch jobs are simulated in a discrete-event simulator, with updates occurring primarily at job arrival, job completion, and capacity change. We model jobs as checkpointable \cite{eisenman2022check}. A job with realized runtime $r$ is divided into fixed-length execution segments of length $\ell_{\mathrm{ckpt}}$, except possibly for the final segment. Each segment inherits the parent job's GPU request, and segment $k+1$ becomes eligible for scheduling immediately after segment $k$ completes.

We consider two queue-ordering rules. Under first-come, first-served (FCFS) with conservative backfilling \cite{4180346}, jobs are ordered by submit time and the scheduler attempts to start the head-of-queue job as early as feasible. If the head job cannot start immediately, the scheduler computes its earliest feasible reservation time. Later jobs may then start out of order only if they can complete before that reserved time, so the head job is never delayed. As a sensitivity case, we also consider a size-based scheduler \cite{harchol2003size} that orders jobs first by requested GPU count and then, among jobs with the same GPU request, by realized runtime, in the spirit of clairvoyant scheduling \cite{motwani1994nonclairvoyant}. We refer to this rule as smallest-work-first (SWF). This rule is not intended as a literal production scheduler; rather, it serves as a size-based sensitivity case for the qualitative robustness of the results. Under both rules, once a job or job segment is started, it occupies GPUs for its realized runtime, or segment runtime in the checkpointed case.

Because the exact production batch-job scheduler is not identifiable from the trace, we do not interpret either modeled rule as a literal reconstruction of the observed system. Instead, we replay the observed batch jobs under each scheduler and compare the resulting aggregate execution dynamics with those in the trace. Batch jobs from all resource groups are processed by the same scheduler. Within this replay exercise, the capacity faced by the unknown production scheduler is not directly observed, so we construct a revealed-capacity proxy to approximate the operating conditions of the original system. Specifically, we compute the daily 99th percentile of busy GPUs and define revealed capacity as the running maximum of these daily 99th-percentile values (Supplementary Figure~9). Using this proxy-capacity timeline, we compare the resulting busy-GPU series and power spectral density with those of the observed trace. Across FCFS with conservative backfilling, the SWF sensitivity case, and an alternative checkpoint length, the replayed execution series reproduce the main concurrency and spectral structure of the trace (Supplementary Figures~10--12). This supports the use of the framework for forward scenario analysis under alternative scheduling rules. In the forward simulation experiments, by contrast, GPU capacity is specified directly by the scenario design.

\textit{Batch--inference co-simulation.}
The inference-side GPU-use and energy measurements are based on H100 GPUs, whereas the batch power traces were collected on V100 GPUs. We therefore adopt H100 as a common hardware baseline for co-simulation. Specifically, batch-side GPU power is multiplied by a constant hardware-adjustment factor so that batch power is expressed on the same H100-based per-GPU scale used for the inference-side measurements. This harmonization is intended to support relative comparisons of variability and ramping across workload compositions, rather than hardware-specific prediction of absolute facility power. Under this assumption, batch and inference workloads share a fixed GPU pool under a shared-GPU policy in which latency-sensitive inference receives priority access to GPUs and batch jobs run on the residual capacity left after inference service \cite{wang2025colocating, chen2024latency}. Depending on the scenario, inference access to the shared GPU pool may be either capped or uncapped. The realized inference allocation is treated as exogenous to the batch scheduler, which operates only on the remaining GPU capacity. Total system power is obtained by summing realized batch and inference power over time.

In the hybrid experiments, total GPU capacity is specified exogenously by the scenario design. We compare scenarios within subsets that hold total GPU capacity and target utilization fixed, using the utilization proxy described in Supplementary Note~2. This provides a common basis for comparing workloads with very different service characteristics, including long batch jobs and short inference requests. Because batch and inference may differ in power per active GPU, power time series are additionally normalized across scenarios. Under this design, differences in realized power profiles are attributable to workload composition and shared-capacity interaction rather than to trivial differences in overall load.

\subsection*{Outcome metrics and normalization}

We characterize the grid-facing load dynamics of hybrid systems primarily in terms of workload composition and utilization. Let $W_{\mathrm{inf}}$ and $W_{\mathrm{batch}}$ denote offered inference and batch GPU work, respectively. We define the inference share as
\[
\frac{W_{\mathrm{inf}}}{W_{\mathrm{inf}}+W_{\mathrm{batch}}}.
\]
Utilization is defined as offered GPU work relative to available GPU capacity described in Supplementary Note~2 and the known scenario capacity in forward simulations.

Aggregate variability is measured by the coefficient of variation of total system power over time. Short-horizon ramping is measured using the absolute change in normalized total power over a specified time horizon $\Delta t$; in the main text we focus on 15-minute ramps, with 1-minute and 5-minute variants reported in the Supplementary Information. Unless otherwise stated, power is normalized by the scenario-specific mean total power so that comparisons reflect temporal structure rather than differences in scale.

\bmhead{Supplementary information}
Supplementary information is available as an ancillary file on arXiv. 

\bmhead{Acknowledgements}
This work was supported in part by the Harvard Belfer–SEAS Faculty Research Fund, and in part by the U.S. Department of Energy (DOE) through the OPEN COG Grid project. We also thank the Harvard Power and AI Initiative for many insightful discussions.

\section*{Declarations}

\begin{itemize}
\item Funding: Not applicable.

\item Conflict of interest/Competing interests: The authors declare the following competing interests. L.X. is a member of the Board of Managers of PJM Interconnection, L.L.C. The views expressed in this paper are the opinions of the authors and do not reflect the views of PJM Interconnection, L.L.C. or its Board of Managers.

\item Ethics approval and consent to participate: Not applicable.

\item Consent for publication: All authors have consented to the publication of this manuscript.

\item Data availability: Batch-job traces with job-level power profiles are openly available at \url{s3://mit-supercloud-dataset/datacenter-challenge/202201/}. Inference-request traces are openly available at \url{https://github.com/Azure/AzurePublicDataset/blob/master/AzureLLMInferenceDataset2024.md}. Because the inference trace does not include direct power measurements, the inference workload model is augmented using serving-performance and energy values from the platform available at \url{https://ml.energy/leaderboard/}. The specific archived values used in this study are provided in the Supplementary Files. 

\item Materials availability: Not applicable.

\item Code availability: The replication package containing the calibration code, processed model artifacts, supporting data, archived experimental outputs, and post-analysis scripts required to reproduce the figures and tables reported in the manuscript is archived at Zenodo \url{https://doi.org/10.5281/zenodo.19463785}. 
The full scenario-generation simulator is currently being privately shared only with the editors and reviewers.


\item Author contributions: S.M. conceived the research, developed the model, performed the simulations and analyses, and prepared the original draft. M.Y. and L.X. supervised the work. All authors discussed and interpreted the results and contributed to the manuscript.
\end{itemize}

\noindent


\bibliography{sn-bibliography}

@misc{lbnl2024datacenters,
  author      = {Shehabi, Arman and Smith, Steven J. and Masanet, Eric and Koomey, Jonathan and Horner, Nathaniel and Shah, Abhishek and Lanzisera, Steven},
  title       = {2024 {United States} Data Center Energy Usage Report},
  year        = {2024},
  url         = {https://eta.lbl.gov/publications/2024-lbnl-data-center-energy-usage-report},
  note        = {Lawrence Berkeley National Laboratory, Berkeley, California. LBNL-2001637. Accessed June 19, 2025}
}

@misc{IEA2024,
  title = { Energy Demand from {AI}},
  author = {{International Energy Agency}},
  year = {2024},
  note = {Accessed July 10, 2025},
  url =  {https://www.iea.org/reports/energy-and-ai/energy-demand-from-ai},
}

@misc{lee2024datacenter,
  author       = {Lee, Victor},
  title        = {{U.S.} Data Center Power Outlook: Balancing Competing Power Consumption Needs},
  year         = {2024},
  url =    {https://www.linkedin.com/pulse/us-data-center-power-outlook-balancing-competing-consumption-lee-iz4pe/},
  note         = {Accessed June 19, 2025}
}

@article{masanet2020recalibrating,
  title={Recalibrating global data center energy-use estimates},
  author={Masanet, Eric and Shehabi, Arman and Lei, Nuoa and Smith, Sarah and Koomey, Jonathan},
  journal={Science},
  volume={367},
  number={6481},
  pages={984--986},
  year={2020},
  publisher={American Association for the Advancement of Science}
}

@article{chen2021incentive,
  title={Incentive-compatible demand response for spatially coupled internet data centers in electricity markets},
  author={Chen, Min and Gao, Ciwei and Shahidehpour, Mohammad and Li, Zuyi},
  journal={IEEE Transactions on Smart Grid},
  volume={12},
  number={4},
  pages={3056--3069},
  year={2021},
  publisher={IEEE}
}

@article{riepin2025spatio,
  title={Spatio-temporal load shifting for truly clean computing},
  author={Riepin, Iegor and Brown, Tom and Zavala, Victor M},
  journal={Advances in Applied Energy},
  volume={17},
  pages={100202},
  year={2025},
  publisher={Elsevier}
}

@inproceedings{acun2023carbon,
  title={Carbon explorer: A holistic framework for designing carbon aware datacenters},
  author={Acun, Bilge and Lee, Benjamin and Kazhamiaka, Fiodar and Maeng, Kiwan and Gupta, Udit and Chakkaravarthy, Manoj and Brooks, David and Wu, Carole-Jean},
  booktitle={Proceedings of the 28th ACM International Conference on Architectural Support for Programming Languages and Operating Systems, Volume 2},
  pages={118--132},
  year={2023}
}

@article{weron2014electricity,
  title={Electricity price forecasting: A review of the state-of-the-art with a look into the future},
  author={Weron, Rafa{\l}},
  journal={International Journal of Forecasting},
  volume={30},
  number={4},
  pages={1030--1081},
  year={2014},
  publisher={Elsevier}
}

@misc{stenclik2021redefining,
  author = {{Redefining Resource Adequacy Task Force}},
  title  = {Redefining Resource Adequacy for Modern Power Systems},
  year   = {2021},
  url    = {https://www.esig.energy/wp-content/uploads/2021/08/ESIG-Redefining-Resource-Adequacy-2021.pdf},
  note   = {Energy Systems Integration Group, Reston, VA. Accessed June 19, 2025}
}

@misc{nerc2010flexibility,
  title     = {Flexibility Requirements and Metrics for Variable Generation: Implications for System Planning Studies},
  author    = {{Integration of Variable Generation Task Force}},
  year      = {2010},
  url       = {https://www.nerc.com/files/ivgtf1-4_final.pdf},
  note   = {North American Electric Reliability Corporation, Atlanta, GA. Accessed June 19, 2025}
}

@misc{gattaciecca2020identifying,
  title={Identifying Effective Demand Response Program Designs for Residential Customers},
  author={Gattaciecca, Julien and Trumbull, Kelly and Krumholz, Samuel and McKanna, Kelley and DeShazo, J. R.},
  year={2020},
  note={California Energy Commission. Publication Number: CEC-500-2020-072. Accessed June 19, 2025},
  url = {https://www.energy.ca.gov/sites/default/files/2021-05/CEC-500-2020-072.pdf}
}

@misc{GHGProtocolScope2_2023,
  author = {{Greenhouse Gas Protocol}},
  title  = {Scope 2 Guidance: An amendment to the {GHG} Protocol Corporate Standard},
  year   = {2023},
  month  = mar,
  url    = {https://ghgprotocol.org/sites/default/files/2023-03/Scope%202%20Guidance.pdf},
  note   = {Convened by the World Resources Institute and the World Business Council for Sustainable Development. Accessed July 10, 2025}
}

@techreport{mural2026aigrid,
  author       = {Mural, Rachel and Pherwani, Dipesh and Gupta, Chaitanya and Yu, Yiqi and Takahashi, Ai and Kim, Dongjoo and Majumder, Subir and Lee, Henry and Yu, Minlan and Xie, Le},
  title        = {{AI}, Data Centers, and the {U.S.} Electric Grid: A Watershed Moment},
  institution  = {Belfer Center for Science and International Affairs},
  year         = {2026},
  month        = feb,
}

@inproceedings{verma2015large,
  title={Large-scale cluster management at {Google} with {Borg}},
  author={Verma, Abhishek and Pedrosa, Luis and Korupolu, Madhukar and Oppenheimer, David and Tune, Eric and Wilkes, John},
  booktitle={Proceedings of the tenth european conference on computer systems},
  pages={1--17},
  year={2015}
}

@inproceedings{yoo2003slurm,
  title={Slurm: Simple {Linux} utility for resource management},
  author={Yoo, Andy B and Jette, Morris A and Grondona, Mark},
  booktitle={Workshop on job scheduling strategies for parallel processing},
  pages={44--60},
  year={2003},
}

@article{burns2016borg,
  title={{Borg, Omega, and Kubernetes}},
  author={Burns, Brendan and Grant, Brian and Oppenheimer, David and Brewer, Eric and Wilkes, John},
  journal={Communications of the ACM},
  volume={59},
  number={5},
  pages={50--57},
  year={2016},
  publisher={ACM New York, NY, USA}
}

@inproceedings{eisenman2022check,
  title={{Check-N-Run}: A checkpointing system for training deep learning recommendation models},
  author={Eisenman, Assaf and Matam, Kiran Kumar and Ingram, Steven and Mudigere, Dheevatsa and Krishnamoorthi, Raghuraman and Nair, Krishnakumar and Smelyanskiy, Misha and Annavaram, Murali},
  booktitle={19th USENIX Symposium on Networked Systems Design and Implementation (NSDI 22)},
  pages={929--943},
  year={2022}
}

@inproceedings{kwon2023efficient,
  title={Efficient Memory Management for Large Language Model Serving with {PagedAttention}},
  author={Kwon, Woosuk and Li, Zhuohan and Zhuang, Siyuan and Sheng, Ying and Zheng, Lianmin and Yu, Cody Hao and Gonzalez, Joseph and Zhang, Hao and Stoica, Ion},
  booktitle={Proceedings of the 29th Symposium on Operating Systems Principles},
  pages={611--626},
  year={2023}
}

@inproceedings{qian2024alibaba,
  title={{Alibaba HPN}: A Data Center Network for Large Language Model Training},
  author={Qian, Kun and Xi, Yongqing and Cao, Jiamin and Gao, Jiaqi and Xu, Yichi and Guan, Yu and Fu, Binzhang and Shi, Xuemei and Zhu, Fangbo and Miao, Rui and others},
  booktitle={Proceedings of the ACM SIGCOMM 2024 Conference},
  pages={691--706},
  year={2024}
}

@inproceedings{wang2025colocating,
  title={Colocating {ML} Inference and Training with Fast {GPU} Memory Handover},
  author={Wang, Jiali and Wang, Yankui and Han, Mingcong and Chen, Rong},
  booktitle={2025 USENIX Annual Technical Conference (USENIX ATC 25)},
  pages={1657--1675},
  year={2025}
}

@inproceedings{chen2024latency,
  title={Latency-guaranteed co-location of inference and training for reducing data center expenses},
  author={Chen, Guoyu and Subramaniyan, Srinivasan and Wang, Xiaorui},
  booktitle={2024 IEEE 44th International Conference on Distributed Computing Systems (ICDCS)},
  pages={473--484},
  year={2024},
}

@inproceedings{10.1145/3620666.3651329,
author = {Patel, Pratyush and Choukse, Esha and Zhang, Chaojie and {Goiri, \'I.} and Warrier, Brijesh and Mahalingam, Nithish and Bianchini, Ricardo},
title = {Characterizing Power Management Opportunities for {LLMs} in the Cloud},
year = {2024},
isbn = {9798400703867},
publisher = {Association for Computing Machinery},
address = {New York, NY, USA},
booktitle = {Proceedings of the 29th ACM International Conference on Architectural Support for Programming Languages and Operating Systems, Volume 3},
pages = {207--222},
numpages = {16},
location = {La Jolla, CA, USA},
series = {ASPLOS '24}
}

@article{choukse2025power,
  title={Power stabilization for {AI} training datacenters},
  author={Choukse, Esha and Warrier, Brijesh and Heath, Scot and Belmont, Luz and Zhao, April and Khan, Hassan Ali and Harry, Brian and Kappel, Matthew and Hewett, Russell J and Datta, Kushal and others},
  journal={arXiv preprint},
  year={2025},
  doi={10.48550/arXiv.2508.14318}
}

@article{stojkovic2024towards,
  title={Towards Greener {LLMs}: Bringing Energy-Efficiency to the Forefront of {LLM} Inference},
  author={Stojkovic, Jovan and Choukse, Esha and Zhang, Chaojie and {Goiri, \'I.} and Torrellas, Josep},
  journal={arXiv preprint},
  year={2024},
  doi={10.48550/arXiv.2403.20306}
}

@article{chung2025ml,
  title={The {ML.ENERGY} Benchmark: Toward Automated Inference Energy Measurement and Optimization},
  author={Chung, Jae-Won and Liu, Jiachen and Ma, Jeff J and Wu, Ruofan and Kweon, Oh Jun and Xia, Yuxuan and Wu, Zhiyu and Chowdhury, Mosharaf},
  journal={arXiv preprint},
  year={2025},
  doi={10.48550/arXiv.2505.06371}
}

@inproceedings{samsi2021supercloud,
  title={The {MIT} {Supercloud} Dataset},
  author={Samsi, Siddharth and Weiss, Matthew L and Bestor, David and Li, Baolin and Jones, Michael and Reuther, Albert and Edelman, Daniel and Arcand, William and Byun, Chansup and Holodnack, John and others},
  booktitle={2021 IEEE High Performance Extreme Computing Conference (HPEC)},
  pages={1--8},
  year={2021},
}

@inproceedings{stojkovic2025dynamollm,
  title={{DynamoLLM}: Designing {LLM} Inference Clusters for Performance and Energy Efficiency},
  author={Stojkovic, Jovan and Zhang, Chaojie and {Goiri, \'I.} and Torrellas, Josep and Choukse, Esha},
  booktitle={2025 IEEE International Symposium on High Performance Computer Architecture (HPCA)},
  pages={1348--1362},
  year={2025},
}

@article{ding2018emission,
  title={Emission-aware stochastic resource planning scheme for data center microgrid considering batch workload scheduling and risk management},
  author={Ding, Zhaohao and Xie, Liye and Lu, Ying and Wang, Peng and Xia, Shiwei},
  journal={IEEE Transactions on Industry Applications},
  volume={54},
  number={6},
  pages={5599--5608},
  year={2018},
}

@article{yu2016distributed,
  title={Distributed real-time energy management in data center microgrids},
  author={Yu, Liang and Jiang, Tao and Zou, Yulong},
  journal={IEEE Transactions on Smart Grid},
  volume={9},
  number={4},
  pages={3748--3762},
  year={2016},
  publisher={IEEE}
}

@article{lublin2003workload,
  title={The workload on parallel supercomputers: modeling the characteristics of rigid jobs},
  author={Lublin, Uri and Feitelson, Dror G},
  journal={Journal of Parallel and Distributed Computing},
  volume={63},
  number={11},
  pages={1105--1122},
  year={2003},
  publisher={Elsevier}
}

@article{ward1963hierarchical,
  title={Hierarchical grouping to optimize an objective function},
  author={Ward, Joe H., Jr.},
  journal={Journal of the American Statistical Association},
  volume={58},
  number={301},
  pages={236--244},
  year={1963},
  publisher={Taylor \& Francis}
}

@book{hilbe2011negative,
  author    = {Hilbe, Joseph M.},
  title     = {Negative Binomial Regression},
  edition   = {2},
  year      = {2011},
  publisher = {Cambridge University Press},
  address   = {Cambridge, UK}
}

@article{aitchison1982statistical,
  title={The statistical analysis of compositional data},
  author={Aitchison, John},
  journal={Journal of the Royal Statistical Society: Series B (Methodological)},
  volume={44},
  number={2},
  pages={139--160},
  year={1982},
  publisher={Wiley Online Library}
}

@book{gelman1995bayesian,
  author    = {Gelman, Andrew and Carlin, John B. and Stern, Hal S. and Rubin, Donald B.},
  title     = {Bayesian Data Analysis},
  year      = {1995},
  publisher = {Chapman and Hall/CRC},
  address   = {New York}
}

@article{katz1987estimation,
  title={Estimation of probabilities from sparse data for the language model component of a speech recognizer},
  author={Katz, Slava},
  journal={IEEE Transactions on Acoustics, Speech, and Signal Processing},
  volume={35},
  number={3},
  pages={400--401},
  year={1987},
  publisher={IEEE}
}

@inproceedings{sheng2024fairness,
  title={Fairness in serving large language models},
  author={Sheng, Ying and Cao, Shiyi and Li, Dacheng and Zhu, Banghua and Li, Zhuohan and Zhuo, Danyang and Gonzalez, Joseph E and Stoica, Ion},
  booktitle={18th USENIX Symposium on Operating Systems Design and Implementation},
  pages={965--988},
  year={2024}
}

@inproceedings{zhong2024distserve,
  title={{DistServe}: Disaggregating prefill and decoding for goodput-optimized large language model serving},
  author={Zhong, Yinmin and Liu, Shengyu and Chen, Junda and Hu, Jianbo and Zhu, Yibo and Liu, Xuanzhe and Jin, Xin and Zhang, Hao},
  booktitle={18th USENIX Symposium on Operating Systems Design and Implementation},
  pages={193--210},
  year={2024}
}

@ARTICLE{4180346,
  author={Tsafrir, Dan and Etsion, Yoav and Feitelson, Dror G.},
  journal={IEEE Transactions on Parallel and Distributed Systems}, 
  title={Backfilling Using System-Generated Predictions Rather than User Runtime Estimates}, 
  year={2007},
  volume={18},
  number={6},
  pages={789--803},}

@article{harchol2003size,
  title={Size-based scheduling to improve web performance},
  author={Harchol-Balter, Mor and Schroeder, Bianca and Bansal, Nikhil and Agrawal, Mukesh},
  journal={ACM Transactions on Computer Systems},
  volume={21},
  number={2},
  pages={207--233},
  year={2003},
  publisher={ACM New York, NY, USA}
}

@article{motwani1994nonclairvoyant,
  title={Nonclairvoyant scheduling},
  author={Motwani, Rajeev and Phillips, Steven and Torng, Eric},
  journal={Theoretical Computer Science},
  volume={130},
  number={1},
  pages={17--47},
  year={1994},
  publisher={Elsevier}
}

\end{document}